\documentclass[prd,amsmath,amssymb,preprintnumbers,showpacs,showkeys,twocolumn]{revtex4}
\newcommand{\ip}[2]{{\langle #1\mid #2\rangle}}

\usepackage{amsmath,amssymb,mathrsfs}

\begin{document}
\title{Quantum energy inequalities in two dimensions}
\author{Christopher J. Fewster}
\email{cjf3@york.ac.uk}
\affiliation{Department of Mathematics, University of York, Heslington, York, YO10 5DD, UK}
\date{\today}
\begin{abstract} 
Quantum energy inequalities (QEIs) were established by Flanagan for the
massless scalar field on
two-dimensional Lorentzian spacetimes globally conformal to Minkowski space.
We extend his result to all two-dimensional globally hyperbolic
Lorentzian spacetimes and use it to show that flat spacetime
QEIs give a good approximation to the curved
spacetime results on sampling timescales short in comparison with
natural geometric scales. This is relevant to the application of QEIs to
constrain exotic spacetime metrics.
\end{abstract}
\pacs{04.62.+v}
\keywords{Energy conditions, Quantum inequalities}
\maketitle
%


Classically, the massless (minimally coupled) free scalar field obeys the weak energy
condition: it displays a nonnegative energy density to all observers
at all points in spacetime. Its quantised sibling is quite different, however,
admitting unboundedly negative energy densities at individual spacetime
points. Violations of the energy conditions are cause for concern, and
a considerable effort has been expended, beginning with the work of
Ford~\cite{Ford78}, in trying to understand what
constraints quantum field theory might place on such effects. 
It turns out that averages of the energy density along, for example, timelike
curves obey state-independent lower bounds called quantum inequalities,
or quantum energy inequalities (QEIs). 
In two-dimensional Minkowski space, for example, the massless free field obeys the
QEI bound~\cite{Flanagan97}
\begin{equation}
\int_\gamma \langle T_{ab}(\gamma(\tau))\rangle_\omega u^a u^b G(\tau) d\tau \ge
- \frac{1}{24\pi}\int_{-\infty}^\infty \frac{G'(\tau)^2}{G(\tau)} d\tau
\label{eq:flatQI}
\end{equation}
for all Hadamard states $\omega$,
where $\gamma$ is the worldline of an inertial observer parametrised by
proper time $\tau$ with two-velocity $u^a$, and $G$ is any smooth,
nonnegative sampling function of compact support [i.e., vanishing
outside a compact interval]. The right-hand side is large and negative
if $G$ is tightly peaked, but small if it is broadly spread. Thus
the magnitude and duration are constrained by a relationship reminiscent of
the uncertainty relations: in $d$-dimensional Minkowski space, 
the energy density can drop below $\rho_0<0$ for a time $\tau_0$ only if
$|\rho_0|\tau_0^d<\kappa_d$ for some (small) constant $\kappa_d$ ($\kappa_2=\pi/6$, for example). 

Many exotic spacetimes (wormholes, warp drives, etc) entail violations
of the weak energy condition and it has often been suggested that
quantum fields might provide the necessary distributions of
stress-energy. Quantum energy
inequalities provide a quantitative check on such proposals and have
been used to argue that exotic spacetimes are tightly constrained~\cite{FRworm,FPwarp}.
As no curved spacetime QEIs were available when these references were
written, they made use of flat spacetime
QEIs, and the validity of their conclusions depends on the assumption that quantum fields
in curved spacetimes are subject to the same
restrictions as those in flat spacetimes, at least on sampling
timescales short in comparison with natural geometric scales. We will
refer to this as the `usual assumption': one of our aims is to establish
its general validity in arbitrary two-dimensional globally hyperbolic
spacetimes. 

Our main tool will be the QEI established for the massless scalar field
by Flanagan~\cite{Flanagan02} (based on an earlier 
argument due to Vollick~\cite{Vollick00}). 
As this was originally proved only for those two-dimensional spacetimes which are
globally conformal to Minkowski space, we begin by obtaining a
generalisation to an arbitrary two-dimensional globally hyperbolic spacetime
$(M,g_{ab})$. The same argument applies to spacetimes with boundaries. 

We make use of two observations, the first of which is that
any point $p\in M$ has a `diamond neighbourhood' of the form
$D={\rm int} (J^+(q)\cap J^-(r))$ such that $(D,g_{ab}|_D)$
(considered as a spacetime in its own right) is
globally conformal to the whole of Minkowski space. This may be seen by
introducing null coordinates $(u,v)$ in a neighbourhood of $p$ so that
the metric takes the form $ds^2=e^{2\sigma}\,du\,dv$
for some smooth function $\sigma$. We may assume, without loss of
generality, that this neighbourhood
contains a diamond neighbourhood $D$ of $p$ corresponding to coordinate ranges
$|u|<u_0,|v|<v_0$, say, and by reparametrising $U=\tan(\pi u/(2u_0))$,
$V=\tan(\pi v/(2v_0))$, we see that $(D,g_{ab}|_D)$ is conformal to the
whole of Minkowski space. Furthermore, we may choose a smooth partition
of unity $\chi_\alpha$ on $M$ such that each $\chi_\alpha$ is supported within
some diamond region $D_\alpha$ of the above type (and only finitely many
$\chi_\alpha$ are nonzero at any point of $M$)~\footnote{The existence
of such a partition of unity follows from a paracompactness argument:
first cover $M$ by open diamond regions each of which has compact closure and is
globally conformal to Minkowski space. By paracompactness, there is a
locally finite refinement $\mathcal{O}_\alpha$ of this cover, so that each
$\mathcal{O}_\alpha$ is contained in one of the original diamond regions and
therefore has compact closure. We then choose $\chi_\alpha$ to be a
subordinate partition of unity.}.

Our second observation is that any state of the field on $(M,g_{ab})$
induces a state of the field on each $(D_\alpha,g_{ab}|_{D_\alpha})$
whose $n$-point functions are simply the restrictions to $D_\alpha$ of the
$n$-point functions on $M$. Now the renormalised stress-energy tensor at any
point $p\in D_\alpha$ is obtained
from derivatives of the difference between the two-point function and the Hadamard
parametrix. Since the latter is determined by the local
geometry alone, and therefore independent of whether $p$ is thought of 
as belonging to $M$ or $D_\alpha$, it follows that the induced state has the
same renormalised stress-energy tensor as that on $(M,g_{ab})$. 

Combining these two observations, any average of the renormalised
stress-energy tensor performed within a compact region of $M$ may be
decomposed into a sum of averages performed in finitely many of the
$D_\alpha$, each of which is subject to the QEIs obtained by
Flanagan~\cite{Flanagan02}. As an example, let $\gamma$ be any smooth
timelike curve parametrised by proper time $\tau$, with two-velocity
$u^a$ and acceleration $a^a=u^b\nabla_b u^a$. We follow the conventions of~\cite{Flanagan02}
in which $u^au_a<0$ for timelike $u^a$, and write
$\widetilde{\chi}_\alpha(\tau)=\chi_\alpha(\gamma(\tau))$. 
Then for any Hadamard state $\omega$ of
the field on $(M,g_{ab})$ and any nonnegative, smooth, compactly
supported sampling function $G$, we have
\begin{widetext}
\begin{eqnarray}
\int_\gamma \langle T_{ab}(\gamma(\tau))\rangle_\omega u^a u^b G(\tau) d\tau
&=&\sum_\alpha \int_\gamma \langle T_{ab}(\gamma(\tau))\rangle_\omega u^a u^b
G(\tau)\widetilde{\chi}_\alpha(\tau) 
d\tau \nonumber\\
&\ge& 
- \frac{1}{24\pi}\sum_\alpha\int_{-\infty}^\infty 
\left[\frac{(\widetilde{\chi}_\alpha G)'(\tau)^2}{\widetilde{\chi}_\alpha(\tau)G(\tau)} 
+\widetilde{\chi}_\alpha(\tau)G(\tau) \left(a^a a_a + R\right)\right] d\tau
\nonumber\\
&=& - \frac{1}{24\pi}\int_{-\infty}^\infty 
\left[\sum_\alpha\frac{(\widetilde{\chi}_\alpha G)'(\tau)^2}{\widetilde{\chi}_\alpha(\tau)G(\tau)}
+G(\tau) \left(a^a a_a + R\right)\right] d\tau
\label{eq:FFQI}
\end{eqnarray}
\end{widetext}
where we have used the decomposition mentioned above, and applied
the QEI Eq.~(1.7) of~\cite{Flanagan02} to the sampling function
$\widetilde{\chi}_\alpha(\tau)G(\tau)$ in each $D_\alpha$~\footnote{The integrand on the
right-hand side of Eq.~\eqref{eq:FQI} is taken
to vanish outside the support of $G$. To convert Eq.~\eqref{eq:FQI} to a signature in
which $u^au_a>0$ for timelike $u^a$, it is necessary only to reverse the sign
of $a^aa_a$.}. Note that only finitely many of
the summands are nonzero, so the right-hand side is finite. In exactly
the same way, we may extend other QEIs obtained in~\cite{Flanagan02},
which involve averages along spacelike or null curves. The same would
apply to spacetime-averaged QEIs of the type considered in~\cite{Flanagan97}.

The above argument significantly extends the range of applications of 
QEIs in two dimensions. In particular, it allows the consideration of the
Kruskal extension of the two-dimensional black hole spacetime
\begin{equation}
ds^2 = -\frac{32M^3e^{-r/(2M)}}{r}dUdV
\end{equation}
where $r=r(U,V)$ is defined implicitly by 
\begin{equation}
\left(1-\frac{r}{2M}\right)e^{r/(2M)} = UV \,,
\end{equation}
and $U$ and $V$ are restricted so that $r(U,V)>0$, i.e., $UV<1$. The horizon is
located at $r=2M$, or equivalently $UV=0$, and exterior Schwarzschild is
the region $U<0$, $V>0$. Although not globally
conformal to Minkowski space, this spacetime can be covered by
diamond neighbourhoods, each of which is globally conformal to Minkowski
space. The only geometrical constraint on these neighbourhoods is that
they should remain inside the physical region $UV<1$. The QEI bounds will
become quite weak for trajectories which approach the singularity, owing to
the divergence of the Ricci scalar $R=4M/r^3$. As noted by 
Flanagan~\cite{Flanagan02}, QEIs along
worldlines which remain close to, but outside, the horizon will also be
rather weak due to their large acceleration. For example, a worldline
with constant $r>2M$ has $a^aa_a = M^2/(r^3(r-2M))$. However, as Roman
has emphasised~\cite{Roman04} (and {\em pace}~\cite{Hayward}) this does not mean that there are no QEI
constraints near the horizon~\footnote{As is evident from
Eq.~\eqref{eq:FFQI}, QEIs along highly accelerated trajectories give
weak bounds even in Minkowski space.}. Indeed, in the case $M\gg 1$, it is clear that QEIs along the
worldline of a freely falling observer passing through the horizon 
differ little from the flat space results for averages near the horizon
(note $R=1/(2M^2)$ at $r=2M$). Of course the averaging must be
completed well before the singularity is reached, which sets an upper
limit on the proper time available.
For further discussion of these issues see~\cite{Vollick00,Flanagan02,Roman04}. 

We are now in a position to prove the validity of the usual assumption.
Since we are concerned with averaging over small scales, it suffices to
consider a single diamond region $D$ which is globally conformal to
Minkowski space. Suppose a smooth timelike curve
$\gamma$ in $D$ may be parametrised by proper time in the interval
$|\tau|<T$, say. Then Eq.~\eqref{eq:FFQI} reduces to
\begin{multline}
\int_\gamma \langle T_{ab}(\gamma(\tau))\rangle_\omega u^a u^b G(\tau) d\tau
\\ \ge
- \frac{1}{24\pi}\int_{-\infty}^\infty 
\left[\frac{G'(\tau)^2}{G(\tau)} +G(\tau) \left(a^a a_a + R\right)\right] d\tau
\label{eq:FQI}
\end{multline}
for any smooth nonnegative `sampling function' $G$ with compact support
in $(-T,T)$.

We note that the QEI bound consists of two parts: the flat
spacetime result, and correction terms due to the acceleration of the
curve and the scalar curvature of spacetime. As we now show,
the first part will dominate if $G$ is peaked on scales short in
comparison with those set by $R$ and $a^a$. Indeed, 
putting
\begin{equation}
A=\sup_\gamma a^a a_a \qquad{\rm and}\qquad
B=\max\{0,\sup_\gamma R\}\,,
\end{equation}
and replacing $G$ by the scaled version $G_{\tau_0}$ defined
by \begin{equation}
G_{\tau_0}(\tau)=\tau_0^{-1}G(\tau/\tau_0)\,,
\label{eq:scaled}
\end{equation}
Eq.~\eqref{eq:FQI} implies that
\begin{equation}
\int_\gamma \langle T_{ab}(\gamma(\tau))\rangle_\omega u^a u^b G_{\tau_0}(\tau) d\tau \ge
-\frac{A+B}{24\pi}-\frac{C}{24\pi\tau_0^2}\,,
\label{eq:FQIs}
\end{equation}
where the constant $C$ is given in terms of the `unscaled' sampling
function as
\begin{equation}
C =\int_{-\infty}^\infty G'(\tau)^2/G(\tau)\, d\tau\,.
\end{equation} 
(We have also
used the fact that $a^a a_a\ge 0$.)
It is easy to find examples of $G$ supported within an interval of unit proper time
with $C$ of the order of $40$ (the minimum
value is $4\pi^2$~\footnote{Set $G(\tau)=2\sin^2(\tau\pi)$ for $0<\tau<1$ and
zero elsewhere, which attains $C=4\pi^2$. Although this function is only $C^1$, there exist smooth $G$ with
values of $C$ arbitrarily close to this value.}).
Accordingly, if 
\begin{equation}
\tau_0\lesssim 10^{-3}\min\{A^{-1/2},B^{-1/2}\}
\end{equation} 
and $\tau_0<2T$ (i.e., sampling occurs within $D$) the second term in Eq.~\eqref{eq:FQIs}
dominates over the first by a factor of around $10$ and the flat space result may be safely
utilised, certainly for the order-of-magnitude considerations required in~\cite{FRworm,FPwarp}.

We have therefore justified the usual assumption for the massless scalar
field in two-dimensions. Three geometric
scales are relevant: the acceleration of the observer, the scalar curvature,
and the maximum size (as measured by $T$) 
of diamond neighbourhood globally conformal to the whole of Minkowski
space. The last of these becomes relevant when the spacetime contains
boundaries or singularities (cf.~\cite{FPR}). 

To conclude, let us consider the status of the usual assumption for
dimensions other than two and/or massive fields. 
Since refs.~\cite{FRworm,FPwarp} were written, QEIs have been
established in curved spacetimes and such results exist for the free
scalar~\cite{FordPfenning98,FTeo,AGWQI,Flanagan02}, Dirac~\cite{Vollick00,FVdirac}, Maxwell and Proca
fields~\cite{Pfenning_em,FewsterPfenning} in various levels of
generality, including very general results. With the exception
of~\cite{Vollick00,Flanagan02} (and those discussed above), these bounds have been ``difference'' QEIs:
namely, the quantity bounded is the difference between the energy
density in the state of interest and that in a reference state. 
For example, in~\cite{AGWQI} a QEI was obtained for 
the scalar field in an arbitrary globally hyperbolic spacetime
$(M,g_{ab})$ for sampling along any smooth timelike curve $\gamma$,
which took the form
\begin{multline}
\int_\gamma \left[\langle T_{ab}(\gamma(\tau))\rangle_\omega 
-\langle T_{ab}(\gamma(\tau))\rangle_{\omega_0}\right]  u^a u^b G(\tau)
d\tau \\
\ge -{\cal Q}[M,g_{ab},\gamma,\omega_0,G]
\end{multline}
where $\omega_0$ is an (arbitrary, but fixed) reference Hadamard state.
This bound holds for arbitrary Hadamard states $\omega$ and any $G$ of the
form $G(\tau)=g(\tau)^2$ with $g$ real-valued, smooth
and compactly supported; an explicit formula for ${\cal Q}$ can be
given~\cite{AGWQI}. Now it is easy to see that an ``absolute'' QEI
follows immediately, simply by
correcting the ``difference'' bound by the renormalised energy density of the
reference state:
\begin{multline}
\int_\gamma \langle T_{ab}(\gamma(\tau))\rangle_\omega 
u^a u^b G(\tau)
d\tau \\
\ge -{\cal Q}[M,g_{ab},\gamma,\omega_0,G] \\ +
\int_\gamma \langle T_{ab}(\gamma(\tau))\rangle_{\omega_0}u^a u^b G(\tau)\,d\tau\,.
\end{multline}
Of course, the problem is that these expressions depend on the
reference state, and in a general spacetime it is not usually possible to write down a closed
form expression for the stress-tensor of any particular Hadamard state. But now
replace $G$ by its scaled version defined by Eq.~\eqref{eq:scaled}. 
The difference QEI bound is expected to scale as 
$\tau_0^{-d}$ in $d$-dimensions, and to approach the corresponding Minkowski space bound
for sufficiently small $\tau_0$. This indeed occurs in examples~\cite{FTeo}
and a general proof is probably not too difficult. On the other hand,
the second term will approach the constant value 
$\langle T_{ab}(\gamma(0))\rangle_{\omega_0}u^a u^b$ as $\tau_0\to 0$ and is therefore
swamped by the first term when $\tau_0$ is small enough.

To establish the usual assumption we would need to quantify
how small is `small enough'. In examples, the difference QEI approaches
the corresponding Minkowski results on timescales short in comparison
with geometric scales, but there remains the problem of the constant
term arising from the reference state. It has not (yet) been ruled out that the
reference state could make an anomalously large
contribution~\footnote{In order to invalidate the usual assumption,
however, it would be necessary to show that {\em all} Hadamard reference
states gave such an anomalous contribution.} in which case the timescale
$\tau_0$ might have to be chosen very much shorter than natural
geometric scales. In this case the QEI bound would be very weak, and
perhaps insufficient to constrain the geometry as in~\cite{FRworm,FPwarp}.
At present there is,
therefore, a small gap in justifying the usual assumption in dimensions
greater than two. However, the results presented here strengthen the
expectation that it can be bridged. 

\begin{acknowledgments}
I thank \'Eanna Flanagan and Tom Roman for useful comments. 
\end{acknowledgments}

\end{document}